# Probing graphene $\chi^{(2)}$ using a gold photon sieve

Michaël Lobet,[†,§,‡], Michaël Sarrazin, [*,†,§,‡], Francesca Cecchet, [†], Nicolas Reckinger, [†,§], Alexandru Vlad[$], Jean-François Colomer[†,§], Dan Lis[*,†,‡]

[†] Research Centre in Physics of Matter and Radiation (PMR), and [§] Research Group on Carbon Nanostructures (CARBONNAGe), University of Namur (UNamur), 61 rue de Bruxelles, B-5000 Namur, Belgium

[$] Institute of Condensed Matter and Nanosciences, Université catholique de Louvain, Place Louis Pasteur 1, B-1348, Louvain-la-Neuve, Belgium

**ABSTRACT:** Nonlinear second harmonic optical activity of graphene covering a gold photon sieve was determined for different polarizations. The photon sieve consists of a subwavelength gold nanohole array placed on glass. It combines the benefits of efficient light trapping and surface plasmon propagation in order to unravel different elements of graphene second-order susceptibility $\chi^{(2)}$. Those elements efficiently contribute to second harmonic generation. In fact, the graphene-coated photon sieve produces a second harmonic intensity at least two orders of magnitude higher compared with a bare, flat gold layer and an order of magnitude coming from the plasmonic effect of the photon sieve; the remaining enhancement arises from the graphene layer itself. The measured second harmonic generation yield, supplemented by semi-analytical computations, provides an original method to constrain the graphene $\chi^{(2)}$ elements. The values obtained are $|d_{31} + d_{33}| \leq 8.1 \times 10^3$ pm²/V and $|d_{15}| \leq 1.4 \times 10^6$ pm²/V for a second harmonic signal at 780 nm. This original method can be applied to any kind of 2D materials covering such a plasmonic structure.

Besides its atomic thickness, graphene has become a new, challenging, playground for many kinds of photonic applications[1-3], due to its outstanding electrical, mechanical and optical properties. For example, its ability to respond to an externally applied electric field makes graphene of special interest for electro-optic purposes. Its unique light absorption and high electric field confinement properties offer new potential - for instance in nonlinear spectroscopies, through efficient frequency conversion, with multiwave mixing mechanisms[4-6].

However, quantification of graphene nonlinear susceptibilities over metallic surfaces is still sparsely reported[7-11]. Quantitative data are of prime necessity to model processes such as frequency conversion in new types of hybrid devices[7] efficiently. In order to mend this gap, this letter proposes an original method combining experimental and numerical results to reconstruct the second-order susceptibility tensor of hybrid systems that are composed of 2D materials covering plasmonic structures. Here, the hybrid system is a photon sieve, i.e. a flat gold film, perforated according to a honeycomb nanohole pattern and coated with a graphene layer[12-16]. As shown hereafter, using a graphene-coated gold photon sieve enhances second harmonic (SH) conversion efficiency by up to two orders of magnitude, compared with bare flat gold, or, similarly, by up to an order of magnitude, compared with a bare gold photon sieve. This enhancement results from the propagation of surface plasmon polaritons (SPP) at the gold/graphene interface.

Indeed, planar graphene in conjunction with metallic nanostructures enables localized electromagnetic hotspots to be created on the graphene sheet, thereby increasing light absorption by flat graphene above the classical value of 2.3%[16-24]. Moreover, it circumvents any momentum mismatch issues occurring with surface plasmon resonance (SPR) when the photon sieve SPP is excited from an optical illumination at normal incidence[4].

Our experimental results, combined with semi-analytical simulations, make it possible to impose strong constraints on the tensor values of graphene $\chi^{(2)}$. To the best of our knowledge, this is the first time that such an inverse scattering problem method, i.e. a reconstruction of the nonlinear $\chi^{(2)}$ tensor elements of the graphene/sieve from the second harmonic scattered field, has been used to determine nonlinear optical properties of 2D materials.

The gold photon sieve (Figure 1a,b) was fabricated by using colloidal nanosphere lithography. A previous study details the fabrication method extensively [16]. The presence of defect-free single-layer graphene covering the sieve was verified by Raman spectroscopy (Figure 1c). Additional fabrication details and characterization of potential defects can be found in the supporting information (SI). The resulting sieve is a 25-nm-thick gold film perforated by a hexagonal array of holes, with a hole diameter d = 405 nm and a grating parameter $a_0$ = 980 nm (Figure 1a,b). Consequently, the optical resonances of the photon sieve occur in the near infrared regime and depend on the light's incident angle $\theta$. At $\theta = 18°$, the maximum absorption peak occurs at 1560 nm: this matches the regular laser-based optical communication band (Figure 1d). The optical measurements were performed with a Ti:sapphire fiber laser (Toptica - Femtoferb) operating at 1560 nm and delivering pulses of 50 fs at a repetition rate of 100 MHz. As shown in Figure 1e, the laser was focused down to a 100 μm spot on the sample. The average power on the sample was 10 mW (26 MW peak power). It was then filtered to isolate the SH contribution, which was finally captured by a CCD camera (Figure 1f). Additional details about the fabrication method, the angular dependence, the laser source and the setup can be found in the SI.

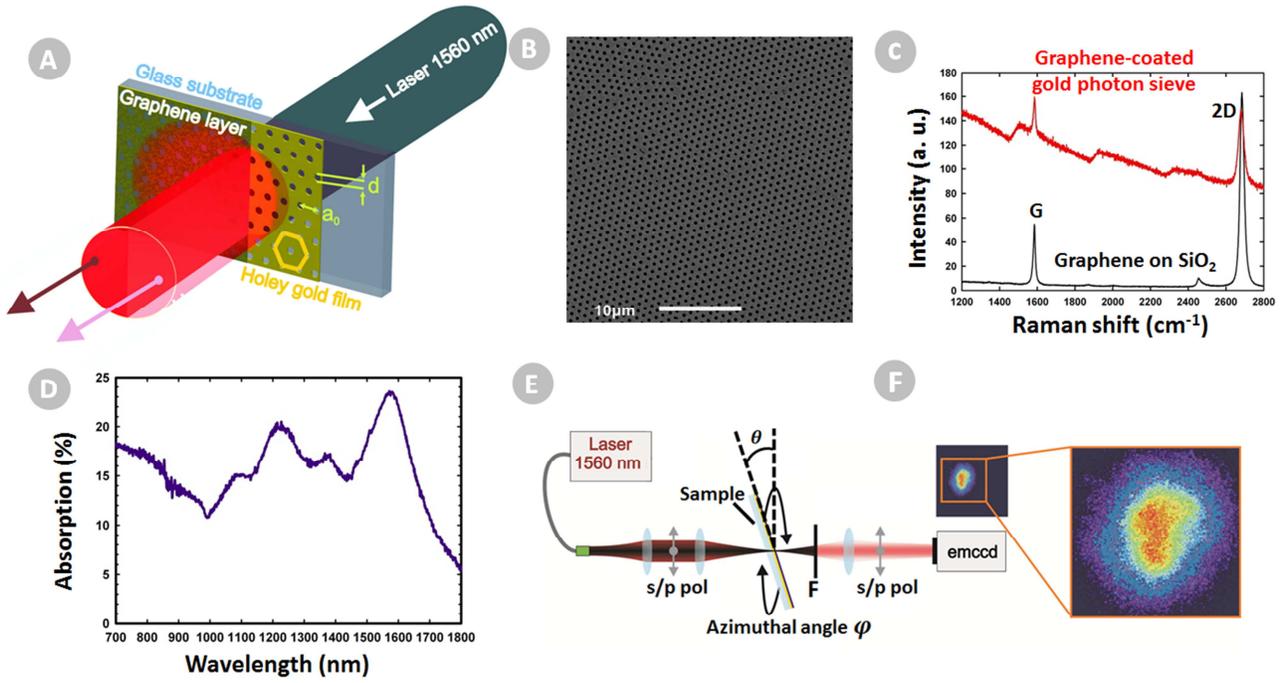

**Figure 1.** (a) Schematic representation of the device and of second harmonic generation (SHG) measurements in transmission-mode. (b) Scanning electron microscopy images of the gold photon sieve. (c) Raman spectrum of the graphene layer on the gold photon sieve and on silicon dioxide. (d) Absorption spectrum of the photon sieve between 700 and 1800 nm at an 18° incident angle. (e) Schematic representation of the SHG spectrometer. (f) Typical SHG image, as collected for the graphene-coated sieve by a CCD camera.

The measurement of SH frequency conversion efficiency has been performed on four samples: I) a bare flat gold layer (25-nm-thick); II) a graphene-coated flat gold layer; III) a bare gold photon sieve; and finally IV) a graphene-coated gold photon sieve. The SHG intensity was measured by scanning the in-plane rotation of the sample over 360° (azimuthal angle $\varphi$), while keeping the incident polar angle fixed at $\theta = 18°$ (Figure 1e). Figure 2 shows the SH polarization maps recorded for SS and PS polarization combinations (e.g.: "PS" stands for P-polarized incident light at ω and S-polarized SHG signal at 2ω). Other polarization maps and SHG intensities can be found in the SI. The SHG maps of the flat gold film are marked as green triangles. Since the gold surface is azimuthally isotropic, the SH intensity is independent of the sample rotation[25,33]. Interestingly, the same SHG response (i.e. mapping profile and SHG intensity) was obtained for the graphene-coated flat gold sample, indicating that graphene mainly does not contribute to the SH signal. For this reason, we will not consider this sample any more in the rest of the discussion. However, it is worth mentioning that an isotropic SHG contribution has been reported for a single-layer graphene supported on oxidized silicon[26]. The SHG emission from the photon sieve is depicted with orange squares in Figure 2. The measured intensity is larger than that of bare flat gold, and shows specific rotational symmetry. Photon sieves are known to concentrate light and increase the electric field locally at their surface. This results from the fact that most of the diffraction orders are evanescent and, consequently, allow surface plasmon polariton (SPP) excitation[13]. As SHG intensity scales with the square of the electric field, it is enhanced accordingly. Moreover, the dependency of the intensity pattern on the azimuthal angle is a function of the array geometry. A hexagonal photon sieve induces 6-fold and 12-fold contributions in the intensity pattern[10,27], as will be discussed later, using a semi-analytical model. SHG maps of the graphene-coated gold sieve are shown as blue circles. The SH intensity is much higher: up to a factor 43 compared with the bare gold photon sieve, and up to 276 compared with flat gold for SS polarization (Figure 2a). For the graphene-coated gold photon sieve, the azimuthal symmetry is less dominant.

The SHG intensities, conversion efficiencies and effective scalar nonlinear susceptibilities $\chi_{eff}^{(2)}$ (see SI for more details) extracted from the measurements are listed in Table 1, in order to facilitate their discussion. SHG intensities (black lines, Figure 2) have been derived by fitting the SHG maps using:

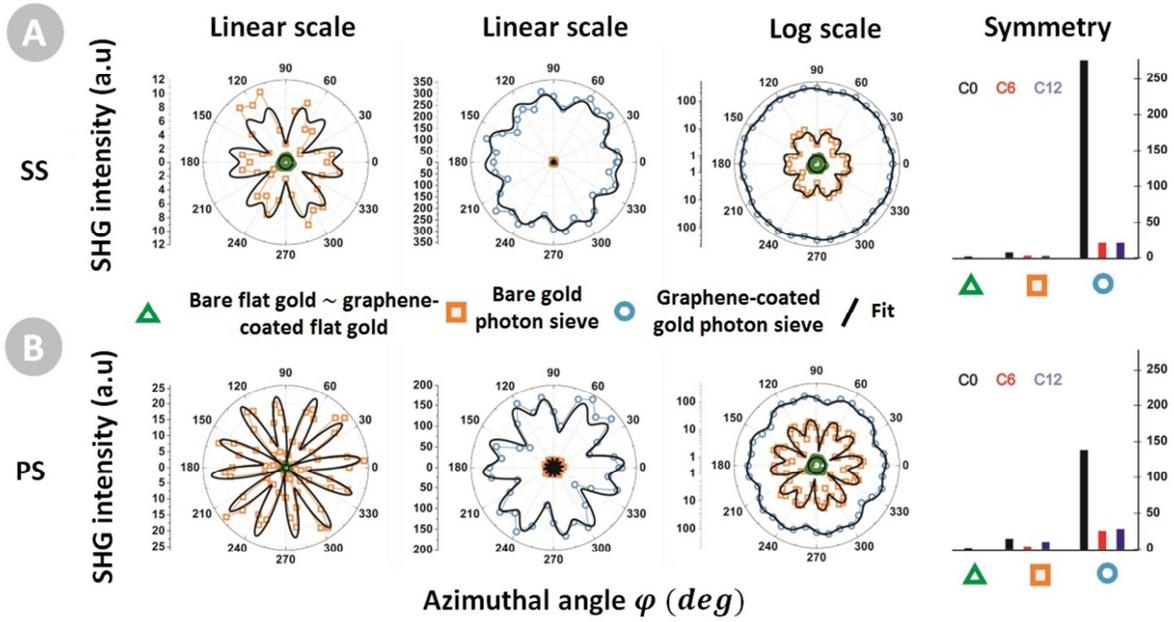

**Figure 2.** Polarization maps of SHG intensity as a function of the azimuthal angle $\varphi$ for bare flat gold and graphene-coated flat gold (green triangles), bare gold photon sieve (orange squares) and graphene-coated gold photon sieve (blue circles) (a) for SS polarization and (b) PS polarization in linear and logarithmic scales. Black lines correspond to fitted values obtained using equation 1. The right panel shows SH intensities decomposed according to their $C_0$, $C_6$, and $C_{12}$ contributions (eq. 1).

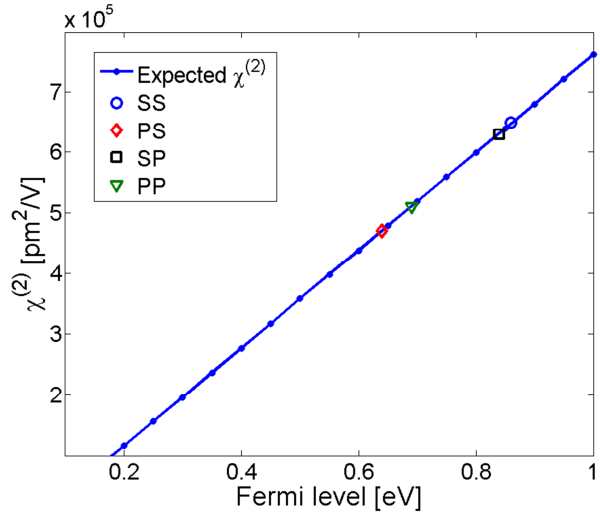

**Figure 3.** Evolution (blue line) of the expected graphene effective second-order susceptibility $\chi^{(2)}_{\text{eff}}$ versus Fermi level $E_f$ as obtained from eq. 2. Blue circle, red rhombus, black square and green triangle correspond to the second-order susceptibility values retrieved from the experimental data, as obtained from SS, PS, SP and PP polarization sets, respectively.

$$I_{SHG}(\varphi) = C_0 + C_6 \cos(6\varphi) + C_{12}\cos(12(\varphi + \varphi_0)) \quad (1)$$

where $\varphi_0$ is the original azimuthal angle, and $C_0$, $C_6$ and $C_{12}$ are the isotropic, the 6-fold and 12-fold contributions to SH intensity, respectively. All the experimental intensities have been compared and normalized to the bare flat gold intensity in SS polarization (Table 1).

Knowing the incident and collected number of photons, an estimated conversion efficiency of the different samples was calculated. Background correction, scaling by the CCD camera gain and quantum efficiency have been taken into consideration (Table 1). The resulting absolute conversion efficiency is in the range of $10^{-14}$. This quite low value can be explained by the large laser spot (100 μm), used in order to prevent any thermal damage and instability. The effective second order $\chi^{(2)}_{\text{eff}}$ values derived from the conversion efficiency are listed in Table 1. The chosen units for $\chi^{(2)}_{\text{eff}}$ are pm²/V (instead of pm/V), since we are interested in the intrinsic surface SHG signal. Indeed, gold SHG response results from a break in symmetry, which is mainly a surface effect. Moreover, graphene is a monoatomic 2D layer, for which only a surface $\chi^{(2)}$ makes sense. The large SH yield enhancement following the graphene coating has to be attributed to the intrinsic second-order susceptibility $\chi^{(2)}$ of the graphene layer itself, as revealed by numerical calculations.

**Table 1. SHG intensities for different samples and different polarization states**

| | | SHG Intensity (normalized) | | | | |
|---|---|---|---|---|---|---|
| Polarization | Sample | $C_0$ contribution | $C_6$ contribution | $C_{12}$ contribution | Conversion efficiency | $\chi^{(2)}_{\text{eff}}$ (pm²/V) |
| SS | Bare flat gold | 1.0 | 0.0 | 0.0 | $7.0 \times 10^{-15}$ | $2.8 \times 10^3$ |
| SS | Bare gold photon sieve | 6.4 | 2.0 | 1.6 | $4.5 \times 10^{-14}$ | $9.2 \times 10^4$ |
| SS | Graphene-coated photon sieve | 276.0 | 20.0 | 20.0 | $1.9 \times 10^{-12}$ | $6.5 \times 10^5$ |
| PS | Bare flat gold | 1.0 | 0.0 | 0.0 | $7.0 \times 10^{-15}$ | $2.8 \times 10^3$ |
| PS | Bare gold photon sieve | 13.6 | 2.8 | 9.2 | $9.5 \times 10^{-14}$ | $1.4 \times 10^5$ |
| PS | Graphene-coated photon sieve | 138.0 | 24.8 | 27.2 | $9.7 \times 10^{-13}$ | $4.7 \times 10^5$ |

**Table 1.** $C_0$, $C_6$ and $C_{12}$ contributions to SHG intensities for the different samples, obtained after fitting the measured responses and normalized to the bare flat gold layer; SH conversion efficiencies and second-order susceptibilities for SS and PS polarization. See SI for calculation details.

The presence of the graphene coating boosts the SH signal 10 to 40 times, meaning that the overall contribution of graphene constitutes at least 90% of the entire signal. The obtained $\chi^{(2)}_{\text{eff}}$ values for the graphene-coated gold photon sieve are 200 times higher than for the bare flat gold sample; as deduced from the SS polarization set, the bare flat gold $\chi^{(2)}_{\text{eff}}$ is $2.8 \times 10^3$ pm²/V, while it reaches $6.5 \times 10^5$ pm²/V for the graphene-coated gold photon sieve (see Table 1). Note that the effective graphene $\chi^{(2)}_{\text{eff}}$ on a gold photon sieve, when normalized relative to a typical graphene thickness (0.3 nm), becomes $2.2 \times 10^3$ pm/V, which is 500 times higher than a typical nonlinear crystal (e.g. barium borate $\chi^{(2)}_{\text{eff}}$ is 4.4 pm/V)[11,30].

The high order of magnitude of the effective second-order $\chi^{(2)}_{\text{eff}}$ of graphene, evidenced by the experimental results, can be retrieved by using a rough theoretical estimate. As a first approximation, a simple classical model gives the expected magnitude of the second-order response[33]:

$$\chi^{(2)}_{\text{eff}} \sim \frac{e}{4m_e \omega^2}(\varepsilon(\omega) - 1). \quad (2)$$

The electric permittivity of graphene $\varepsilon(\omega)$ can be calculated using the well-known Kubo formula for graphene's conductivity[34,35] including intraband and interband terms (see the SI). This model is frequency-, temperature-, and charge-density-dependent. At room temperature and at 1560 nm, the estimated effective second-order $\chi^{(2)}_{\text{eff}}$ of graphene depends on the Fermi level as shown in Figure 3. From this figure, it can be seen that the order of magnitude of the graphene second-order susceptibility, as given by eq. 2, closely matches the experimental values (Table 1) obtained from the four measured polarization sets (SS, PS, SP, PP), provided the Fermi level $E_f$ comprised between 0.65 and 0.85 eV. Such values are consistent with those expected for graphene on gold[36-38]. As an illustration, for a Fermi level of 0.75 eV, the corresponding $\chi^{(2)}_{\text{eff}}$ value corresponds to $5.7 \times 10^5$ pm²/V, which is the mean value of the experimental data. This result further confirms that the high second-order susceptibility of graphene is the main contribution to the SH yield enhancement.

Using a rigorous coupled wave analysis (RCWA) method[13,28,29] clearly demonstrates that the presence of graphene does not modify the electromagnetic field at the photon sieve surface significantly (Figure 4). The electromagnetic field response is only weakly altered, as exemplified by Figures 4a and 4b for the bare sample and the graphene-coated sample, respectively. Furthermore, the weak influence of the graphene coating on the electromagnetic field at the photon sieve surface can be estimated by examining the square modulus of the electric field, integrated in the lateral $x$ and $y$ directions, versus the depth $z$ of the whole sample (Figure 4c). Fermi levels compatible with the above analysis are used. Although variation is observed at the graphene/gold photon sieve interface, this variation is too small to explain the enhanced SHG signal, even if a charge distribution is taken into account. Because of the graphene coating, the electromagnetic field of the gold photon sieve does not show any significant amplification, regardless of the Fermi level.

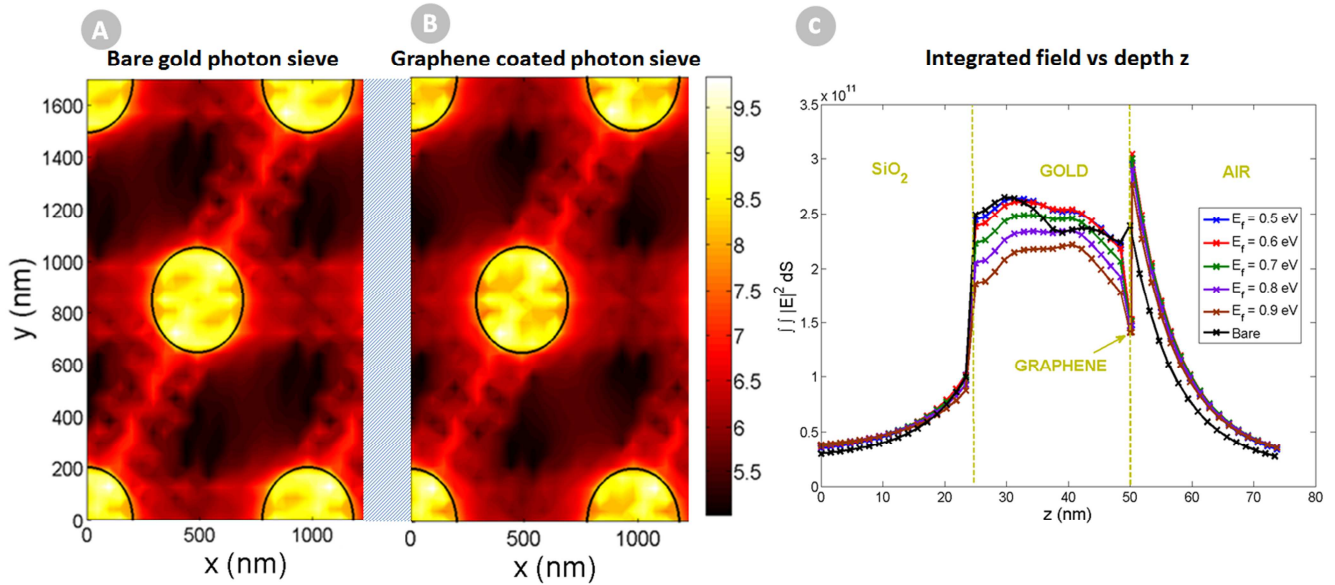

**Figure 4.** Field maps ($|E|^2$) of (a) the bare gold photon sieve at the air-photon sieve interface and (b) of the graphene-coated photon sieve in the graphene layer ($E_f$ = 0.7 eV). Both field maps have a logarithmic scale. (c) Integrated square modulus of the electric field versus depth z for different Fermi level values.

It is important to highlight that those photon sieve properties which allow strong light absorption and the propagation of long range SPP at the gold/graphene interface also enable us to probe new elements of the graphene $\chi^{(2)}$ tensor that were silent in the graphene-coated flat gold hybrid structure. This is why graphene on flat gold does not contribute to the SHG signal, while graphene on a gold photon sieve induces an effective $\chi^{(2)}_{\text{eff}}$ as high as $6.5 \times 10^5$ pm$^2$/V. To further quantify the amplitude of the $\chi^{(2)}$ tensor elements of graphene on a gold photon sieve, both the SHG intensity and azimuthal symmetry patterns have to be considered. As deduced from Figure 2 and summarized in Table 1, the bare flat gold sample (or similarly the graphene-coated one) displays a weak intensity with a unique $C_0$ symmetry. However, the hexagonal lattice of the gold photon sieve introduces evident asymmetry, together with an increase in intensity.

In the SS polarization combination, the total anisotropic contribution ($C_6+C_{12}$) equals 3.6, which is 50% less than the $C_0$ one (6.4). In PS polarization, $C_6 + C_{12}$ nearly equals the isotropic $C_0$ contribution (12.0 versus 13.6), with a marked 12-fold pattern. Indeed, the $C_{12}$ component represents about 77% of the whole anisotropic contribution. The increase in the signal intensity for graphene on the gold photon sieve is mainly associated with the $C_0$ contribution (Figure 2). For instance, in the SS polarization set, the anisotropic ($C_6+C_{12}$) signal is enhanced by a factor close to 10, while the $C_0$ contribution increases 40 times. This is not really surprising, since the single-layer graphene is expected to generate an isotropic contribution[26]. The $C_0$ contribution results from multiscattering processes, as explained hereafter.

To retrieve the $\chi^{(2)}$ components of graphene on the gold photon sieve, we developed the following original semi-analytical method. This method can be universally applied to any 2D material lying on such a plasmonic structure. Starting from the above experimental observations, semi-analytical simulations enable us to constrain the amplitude of some components of the second-order susceptibility tensor for graphene on the gold photon sieve. As demonstrated earlier, although gold induces graphene doping, graphene has no significant influence on the electromagnetic field scattered by the photon sieve (Figure 4). Therefore, graphene can be considered a probe of the scattered field in the vicinity of the gold surface. As a consequence, to a first approximation, the second harmonic field (at 2ω) emitted by graphene is calculated analytically, using the scattered field (at ω), as described below. The scattered field is numerically computed. A brief summary of the semi-analytical method would, therefore, be as follows: in the relation linking the second order polarization field $P_{i,2\omega}$ and the scattered field $E_\omega$, i.e., $P_{i,2\omega} = \varepsilon_0 \chi^{(2)}_{ijk} E_{\omega,j} E_{\omega,k}$, the graphene $\chi^{(2)}_{ijk}$ components are adjusted in such a way that the experimental values of $P_{i,2\omega}$ closely match the numerically computed response of the scattered field $E_\omega$.

Since the photon sieve has a periodic structure, diffraction effects have to be taken into account in the calculations. The electric field of the $n^{\text{th}}$ harmonic at frequency $n\omega$ (where $n$ is an integer) is written, using the periodicity of the structure and Bloch's theorem, as:

$$E_{n\omega}(r) = \sum_{g} E_{n\omega,g} e^{i(n k_\parallel + g)\cdot \rho} e^{i\sqrt{n^2 k^2 - |n k_\parallel + g|^2}|z|} \quad (3)$$

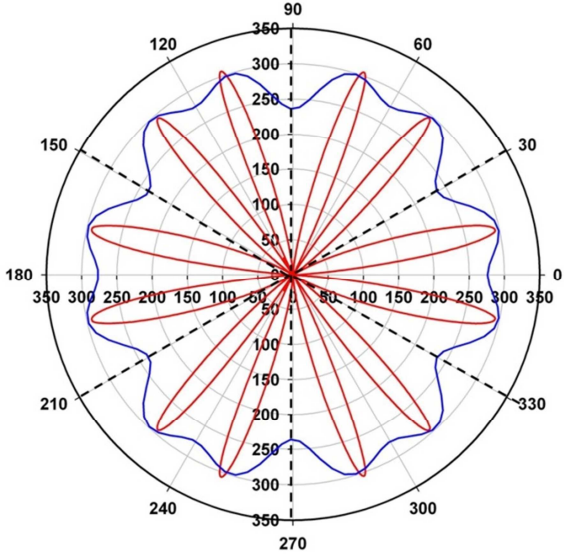

**Figure 5.** Computed (red line) second harmonic signal versus experimental one (blue line) for the SS polarization configuration from the graphene-coated gold photon sieve. The weakly broken $C_{12}$ symmetry as well as the $C_6$ symmetry are reproduced.

where **r** is the position, **g** a reciprocal lattice vector of the photon sieve, **ρ** the position in the (x,y) plane and **k** the incident wave vector. Each term $E_{n\omega,g}$ is numerically computed from the RCWA computational scheme. The polarization fields $P_{n\omega}(r)$ follow the same expression. The electric field $E_{2\omega}(r)$ at frequency $2\omega$ may be unravelled from the second-order polarization field $P_{2\omega}(r)$, as the solution of the wave equation for an electric field $E_{2\omega}(r)$ propagating in a medium of dielectric constant $\varepsilon_{2\omega}(r)$:

$$\nabla \times \nabla \times E_{2\omega}(r) - \frac{(2\omega)^2}{c^2} E_{2\omega}(r) = \left[\frac{(2\omega)^2}{c^2}(\varepsilon_{2\omega}(r) - 1)\right] E_{2\omega}(r) + (2\omega)^2 \mu_0 P_{2\omega}(r). \quad (4)$$

In our approach, the electromagnetic field at $\omega$ is fully propagated in a multiscattering process (thanks to the RCWA method). However, in the following, multiscattering is fairly neglected for the electromagnetic field at $2\omega$. Indeed, the magnitude of the signal at $2\omega$, arising from the specular term, is more important than the details of the azimuthal pattern due to multiscattering. Then, using equation (3), equation (4) can be solved for $E_{2\omega}(r)$ and leads to:

$$E_{2\omega,g} = -\frac{i \mu_0 \omega^2}{\pi |K_z|}\left[\bar{I} - \frac{1}{K^2} K \otimes K\right] P_{2\omega,g} \quad (5)$$

with $K = 2k_\parallel + g + \sqrt{4k^2 - |2k_\parallel + g|^2} e_z$. In the SH experiments carried out in this work, only the first diffraction order (specular) was recorded, i.e. $E_{2\omega,0}$ (see SI for more details). Consequently, only $P_{2\omega,0}$ must be considered. Thus the intensity $I_{2\omega} = (1/2)\varepsilon_0 |E_{2\omega,0}|^2$ can be deduced for each measured polarization, s and p, as follows:

$$I_{2\omega}^{(s)} = \frac{1}{8\pi^2}\frac{\omega^2}{\varepsilon_0 c}\frac{1}{\cos^2\theta}\big(P_{y,2\omega,0}\cos\varphi - P_{x,2\omega,0}\sin\varphi\big)^2 \quad (6)$$

$$I_{2\omega}^{(p)} = \frac{1}{8\pi^2}\frac{\omega^2}{\varepsilon_0 c}\frac{1}{\cos^2\theta}\big(P_{z,2\omega,0}\sin\theta - (P_{x,2\omega,0}\cos\varphi + P_{y,2\omega,0}\sin\varphi)\cos\theta\big)^2 \quad (7)$$

Since, the second-order polarization field is given by $P_{i,2\omega} = \varepsilon_0 \chi^{(2)}_{ijk} E_{\omega,j} E_{\omega,k}$, a straightforward calculation gives the components of $P_{2\omega,0}$:

$$\begin{cases} P_{x,2\omega,0} = 2u\varepsilon_0 d_{15} \\ P_{y,2\omega,0} = 2v\varepsilon_0 d_{15} \\ P_{z,2\omega,0} = \alpha\varepsilon_0 d_{31} + \beta\varepsilon_0 d_{33} \end{cases} \quad (8)$$

with $u = \sum_g \mathrm{Re}\{E_{z,\omega,g} E^*_{x,\omega,g}\}$, $v = \sum_g \mathrm{Re}\{E_{z,\omega,g} E^*_{y,\omega,g}\}$, $\alpha = \sum_g (|E_{x,\omega,g}|^2 + |E_{y,\omega,g}|^2)$, $\beta = \sum_g |E_{z,\omega,g}|^2$, and where $d_{ij}$ corresponds to the components of the convenient reduced nonlinear second-order susceptibility tensor of graphene on gold, with the equivalence relations $d_{31} = d_{32} = \chi^{(2)}_{zyy} = \chi^{(2)}_{zxx}$, $d_{15} = d_{24} = \chi^{(2)}_{xxz} = \chi^{(2)}_{xzx} = \chi^{(2)}_{yzy} = \chi^{(2)}_{yyz}$ and $d_{33} = \chi^{(2)}_{zzz}$ [26,27,31,32].

Since, for centro-symmetric materials $\chi^{(2)} = 0$, the SH signal at $2\omega$ can only be generated from symmetry breaking at the gold photon sieve/graphene/air interfaces; this is precisely where a high density of surface plasmon modes is generated by the incident light $\omega$. Due to symmetry considerations, the only nonzero components of the second order susceptibility tensor are $d_{31} = d_{32}$, $d_{15} = d_{24}$ and $d_{33}$ [7,33].

Once the electromagnetic field amplitude values u, v, α, and β are computed, using the RCWA numerical code, the nonzero tensor elements $d_{15}$, $d_{31}$ and $d_{33}$ are the only parameters needed to retrieve the experimentally measured $I_{2\omega}^{(s)}$ and $I_{2\omega}^{(p)}$ (Eqs. 6 and 7). Those parameters can be fitted using all the experimental data, i.e. for both incident s and p polarizations. Figure 5 shows the retrieved SH map for the SS polarization configuration, as obtained from the semi-analytical procedure above. The quasi-$C_{12}$ (i.e. weakly broken) symmetry as well as the $C_6$ symmetry are reproduced well. The symmetry patterns were also verified when the present model was applied to other incident/transmitted polarization sets (not shown). The origin of the $C_6$ and broken $C_{12}$ symmetry components is consequently explained and reproduced. By contrast, the additional $C_0$ component of each pattern (see Figure 2 and the SI) probably results from a break in symmetry, due to the neglected multi-scattering of the electromagnetic field at $2\omega$ on the photon sieve. Moreover, this accounts for the discrepancy between the experimental and theoretical patterns shown in Figure 5. Whilst this approximation does not change the quantitative estimation of the nonlinear second-order susceptibility tensor, it does prevent us from giving an exact value of tensor elements $d_{15}$, $d_{31}$ and $d_{33}$, and only allow us to set an upper boundary constraint. Numerical results on the electromagnetic field amplitudes indicate that $\alpha \approx \beta$ (Eq. 8), so that a sole constraint can be deduced for $d_{31}$ and $d_{33}$, simultaneously. The so-obtained tensor elements are:

$$\begin{cases} |d_{31} + d_{33}| \leq 8.1 \times 10^3 \text{ pm}^2/\text{V} \\ |d_{15}| \leq 1.4 \times 10^6 \text{ pm}^2/\text{V} \end{cases}$$

where $d_{15}$ and $d_{31} + d_{33}$ have an opposite sign. It should be noted that the above constraints do not exclude $d_{31}$ and $d_{33}$ from being of the same order of magnitude as $d_{15}$, if they have opposite signs. In addition, these constraints on the tensor elements are consistent with the global effective $\chi^{(2)}_{eff}$ evaluated in Table 1, leading to $\chi^{(2)}_{eff} = 5.7\ (\pm 0.8) \times 10^5$ pm²/V (at 95% C.L.), i.e. 14% of relative uncertainty.

In this letter, the nonlinear second harmonic optical activity of monolayer graphene lying on a gold photon sieve was experimentally measured and theoretically modeled. This approach proposes an original method for constraining the $\chi^{(2)}$ tensor elements that combines experimental results, analytical calculations and numerical simulations. The obtained values are $|d_{31} + d_{33}| \leq 8.1 \times 10^3$ pm²/V and $|d_{15}| \leq 1.4 \times 10^6$ pm²/V, for a second harmonic signal at 780 nm. These second-order susceptibility $\chi^{(2)}$ elements contribute to second harmonic generation, due to SPP at the gold photon sieve-graphene interface. The second harmonic intensity is, therefore, enhanced accordingly, by two orders of magnitude. These results demonstrate that, when advantageously coupled to a plasmonic device, several $\chi^{(2)}$ tensor elements of graphene can be involved in generating new frequencies and give highly efficient conversion processes. The retrieval method is general and frequency-independent: it could be applied to any 2D material covering a photon sieve in any wavelength range[39].

## ASSOCIATED CONTENT

**Supporting Information**. Fabrication process, angular polar dependence of SHG intensity, SHG setup, polarization maps and SHG intensities for SP and PP polarizations, effective $\chi^{(2)}_{eff}$ calculation, numerical modelization, approximations. This material is available free of charge via the Internet at http://pubs.acs.org.

## AUTHOR INFORMATION

### Corresponding Authors

* E-mail: michael.sarrazin@unamur.be
* E-mail: dan.lis@unamur.be

### Author Contributions

‡These authors contributed equally, i.e. Michaël Lobet, Michaël Sarrazin and Dan Lis.
The manuscript was written through contributions of all the authors.
All the authors have given their approval to the final version of the manuscript.

## ACKNOWLEDGMENTS

The authors would like to thank Luc Henrard, Karin Derochette and Victoria Welch for careful reading of the work. This research used computing resources of the "Plateforme Technologique de Calcul Intensif (PTCI)" (http://www.ptci.unamur.be) located at the University of Namur, Belgium, which is supported by the F.R.S.-FNRS under convention No. 2.4520.11. The PTCI is a member of the "Consortium des Equipements de Calcul Intensif (CECI)" (http://www.ceci-hpc.be). The research leading to this work received funding from the European Union Seventh Framework Program under grant agreement No 604391 Graphene Flagship. DL, JFC and FC acknowledge the F.R.S.-FNRS for their postdoctoral (DL) and research associate positions (JFC & FC), respectively.

## ABBREVIATIONS

SH second harmonic, SHG second harmonic generation, SPP surface plasmon polariton, SPR surface plasmon resonance, NIR near infrared regime, RCWA rigorous coupled wave analysis.

# Probing graphene $\chi^{(2)}$ using a gold photon sieve

Michaël Lobet,[†,§] Michael Sarrazin,[*,†,§] Francesca Cecchet,[†] Nicolas Reckinger,[†,§] Alexandru Vlad[$], Jean-François Colomer[†,§], Dan Lis[*,†]

[†] Research Centre in Physics of Matter and Radiation (PMR), and [§] Research Group on Carbon Nanostructures (CARBONNAGe), University of Namur (UNamur), 61 rue de Bruxelles, B-5000 Namur, Belgium

[$] Institute of Condensed Matter and Nanosciences, Université catholique de Louvain, Place Louis Pasteur 1, B-1348, Louvain-la-Neuve, Belgium

# Supporting Information

**Table of contents**

- **S1: Fabrication process**

The nanostructured photon sieve (Figure 1a) was fabricated by using colloidal nanosphere lithography. Polystyrene spheres (980 nm in diameter) were deposited on soda-lime glass slides via an interfacial self-assembly protocol [1] and reduced to half the nominal diameter by reactive ion etching using $O_2$ chemistry. After physical vapor deposition of 2 nm of Ti followed by 25 nm of Au, the liftoff was performed using adhesive tape and ultrasonication in dichloromethane. Graphene was grown by atmospheric pressure chemical vapor deposition at 1000 °C on copper foils with methane as hydrocarbon precursor. After synthesis, graphene was transferred onto the holey gold/glass substrate by the polymer-assisted technique. More details regarding the fabrication process can be found in [1].

In summary, the sieve therefore consists of a gold film perforated by a hexagonal array of holes with a hole diameter of $d = 405$ nm and a grating parameter $a_0 = 980$ nm.

Surface characteristics are important for SHG measurements because this technique is intrinsically an interfacial process. Surface deformations of graphene are indeed present, depending on the roughness of the underlying gold film. In addition, pleats related to the wet transfer technique are also present, together with wrinkles resulting from the growth of graphene itself [2,3]. The latter appear due to the difference in thermal dilatation coefficients between graphene and the copper substrate. Moreover, adsorbed water is likely to be trapped between the graphene layer and the gold photon sieve during the transfer of graphene [4,5]. Indeed, the graphene coating of the gold photon sieve involves fishing a PMMA/graphene piece over water. Nevertheless, even if a dry transfer were performed, ambient moisture could also, to a lesser extent, be trapped.

- **S2: Angular polar dependence of SHG intensity**

As suggested by a previous work [1], the efficiency of the holey structure to absorb light depends on the incident angle and on the wavelength. At 1560 nm, which corresponds to the center of the laser source emission spectrum, the gold photon sieve has a maximal absorption at incident angles of 15 to 18° relative to the normal. In all the measurements, an angle of 18° was chosen in order to work under the maximal light absorption condition.

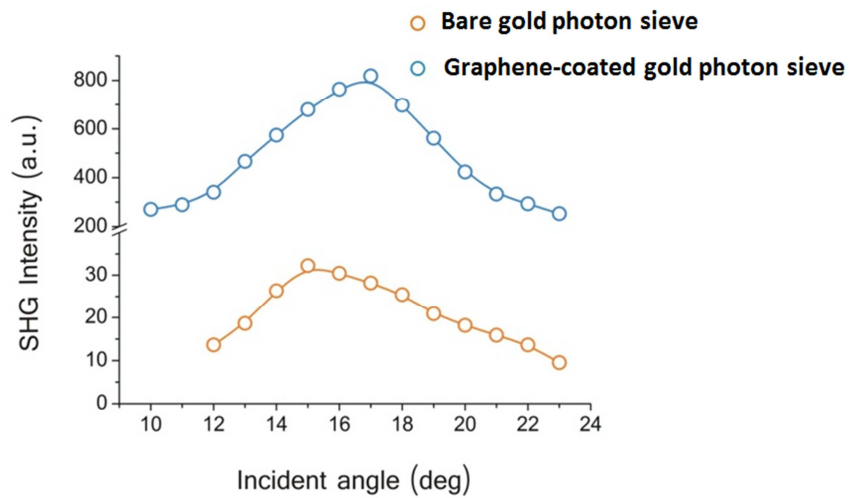

Figure S1: Angular polar dependence of SHG intensity for the bare gold photon sieve and the graphene-coated gold photon sieve.

Given that the sample is tilted by 18° from the normal incidence, the incident light polarization has a single contribution in the sample (y-) plane (s-pol), or is decomposed in both the sample (x-) plane and the sample (z-) normal (p-pol). Identically, either the s-pol or p-pol SHG contribution was collected at the CCD camera.

- **S3: SHG set-up**

The SHG acquisitions were performed with a Ti:Sapphire fiber laser (Toptica - Femtoferb) operating at 1560 nm and delivering pulses of 50 fs at a repetition rate of 100 MHz. The average power at the sample was 10 mW. From those laser characteristics, the peak power at the sample is about 2000 W/s, while the peak power per cm² has been estimated equal to 26 MW/cm² for a circular illuminated area of 100 µm in diameter.

The laser polarization was controlled by a polarizer (Thorlabs – LPNIR) and a half-wave plate (Thorlabs - WPMH1550). Then the fundamental beam was focused at the sample interface with a lens (f = +75 mm), while a short-pass filter was placed prior to the sample to remove light at any harmonic frequency that could have been generated earlier (Thorlabs – FELH1000). The sample is placed on a rotating mount such that both in-plane rotation and light's incident angle variation are allowed. Also, a *x-y* translation stage enables to scan the investigated region on the sample surface. The setup is designed such that it is possible to perform the sample in-plane rotation by keeping the laser spot at the very same location at the interface. The transmitted light went through a colored glass filter (Thorlabs – FGL9) to remove most of the fundamental frequency and was collected by a +25 mm focal length lens. The SH light then went through a polarizer (Thorlabs – LPVis), additional filters (Thorlabs – FL780) and reached the CCD camera (Hamamatsu EM-CCD 9100-13) to produce an image of the interface.

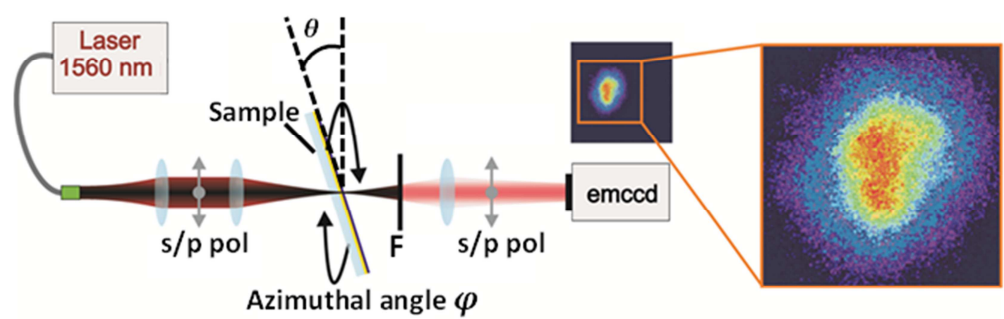

Figure S2: Schematic representation of the SHG setup

Since the experimental conditions (laser power and spot size) have been kept identical, direct intensity and SH yield comparisons can be made. The microscope enlargement has been set to image a sample region of 500 x 500 µm². The laser spot at the sample surface is ~100 µm, making the SHG emission to be at most 100 µm wide (Figure S2, orange frame).

In the recorded spectra, each SHG data point corresponds to a 10 s acquisition time in the amplified EM-CCD mode, where background correction, intensity count and averaging have been applied. Although the spatial resolution is not used for intensity counting, it enables choosing a region of the sample with uniform holes and graphene coverage, corresponding to only one crystallographic domain type. Also, it is required to adjust the sample center such that the in-plane rotation is performed at the very same place (optimized on a sample region presenting some punctual defects), which ensures obtaining a reproducible set of data.

- **S4: Polarization maps and SHG intensities for SP and PP polarizations**

Similar to the full paper, polarization maps and SHG intensities have been recorded for SP and PP polarizations. Figure S3 and Table S1 present those results.

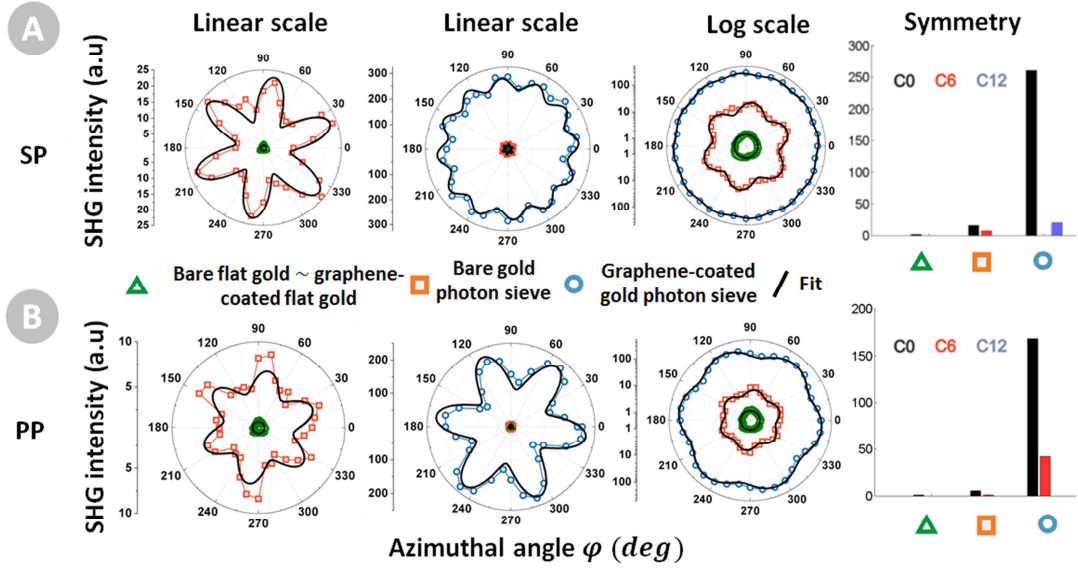

Figure S3: Polarization maps of the SHG intensity as a function of the azimuthal angle $\varphi$ for bare flat gold and graphene-coated flat gold (green triangles), bare gold photon sieve (orange squares) and the graphene-coated gold photon sieve (blue circles) (a) for SP polarization and (b) PP polarization in linear and logarithmic scales. The black lines correspond to fitted values using eq. 1. The right panel shows the SH intensities in logarithmic scale decomposed according to the $C_0$, $C_6$, and $C_{12}$ contributions (eq. 1).

**Table S1. SHG intensities for different samples and different polarization states**

| | | SHG Intensity (normalized) | | | | |
|---|---|---|---|---|---|---|
| Polarization | Sample | $C_0$ contribution | $C_6$ contribution | $C_{12}$ contribution | Conversion efficiency | $\chi^{(2)}_{\text{eff}}$ (pm²/V) |
| SP | Bare flat gold | 1.4 | 0.0 | 0.0 | $9.8 \times 10^{-15}$ | $3.4 \times 10^3$ |
| SP | Bare gold photon sieve | 16.0 | 6.8 | 0.0 | $1.1 \times 10^{-13}$ | $1.5 \times 10^5$ |
| SP | Graphene-coated photon sieve | 260.0 | 0.0 | 20.0 | $1.8 \times 10^{-12}$ | $6.3 \times 10^5$ |
| PP | Bare flat gold | 1.2 | 0.0 | 0.0 | $8.4 \times 10^{-15}$ | $3.1 \times 10^3$ |
| PP | Bare gold photon sieve | 5.6 | 1.2 | 0.0 | $3.9 \times 10^{-14}$ | $8.6 \times 10^4$ |
| PP | Graphene-coated photon sieve | 168.4 | 42.0 | 0.0 | $1.2 \times 10^{-12}$ | $5.1 \times 10^5$ |

**Table 1**. C0, C6, C12 contributions to the SHG intensities for the different samples obtained after fitting the measured responses and normalized to the gold layer, SH conversion efficiencies and second-order susceptibilities for SP and PP polarizations.

- **S5: Effective $\chi^{(2)}$ calculation**

This effective approach is usual in SHG optics. Indeed, while the nonlinear susceptibility tensor links the incident electric vector field $\vec{E_\omega}$ to the SH polarization $\vec{P_{2\omega}}$ through:

$$P_{i,2\omega} = \varepsilon_0 \chi^{(2)}_{ijk} E_{\omega,j} E_{\omega,k},$$

the effective $\chi_{eff}^{(2)}$ is a scalar quantity which links the amplitude of the incident electric vector field $\vec{E_\omega}$ to the amplitude of the SH polarization $\vec{P_{2\omega}}$ through:

$$\left\|\vec{P_{2\omega}}\right\| = \chi_{eff}^{(2)} \left\|\vec{E_\omega}\right\|^2.$$

Since, in the dipolar approximation, the second-order dielectric susceptibility tensor equals zero in bulk centro-symmetric materials, only the interface is supposed to contribute to the SHG generation.

To give an estimate of the effective second-order susceptibility of the different samples, in pm²/V, the following formula was used [6]:

$$\chi_{eff}^{(2)} t = \left(\frac{\varepsilon_0 n_\omega^2 n_{2\omega} c \lambda_\omega^2 S I_{2\omega}}{2\pi^2 I_\omega^2}\right)^{1/2}$$

where $n_{\omega(2\omega)}$ is the refractive index at frequency $\omega$ $(2\omega)$, $\lambda_\omega$ the wavelength at frequency $\omega$, $S$ the cross-section of the illumination spot, $I_{\omega(2\omega)}$ the intensity of the spot at frequency $\omega(2\omega)$ and $t$ the thickness contributing to SHG signal.

- **S6: Numerical modelization**

Rigorous Coupled Wave Analysis is a numerical method traditionally used when the system is stratified and present a periodicity of the refractive index in the lateral directions. This is the case in the present system since the gold photon sieve possesses a lateral periodicity with air inclusions, the other optical media being homogeneous in all directions (air, SiO2 and graphene). The numerical code uses spatial Fourier expansions of the dielectric function for each layer of the structure. The electric and magnetic fields are developed using the same Fourier basis and their Fourier components are propagated throughout the structure by applying electromagnetic boundary conditions at the layer interfaces. The number of plane waves used in the Fourier expansions is the critical convergence parameter. In the present case, $17 \times 17$ plane waves were used to reach numerical convergence within reasonable computation times.

Since the method mainly solves Maxwell's equations by applying the boundary conditions at the layer interfaces, graphene can be accurately modelled by a homogenous layer of thickness 0.34 nm. The reflection, transmission, and absorption coefficients and the field maps are insensitive to the thickness of the layer since it is deeply subwavelength.

The relative difference between the thicknesses present in the photon sieve consequently does not play any role and can be used as it is. However, for numerical accuracy reasons, the gold layer is divided using 16 sublayers and graphene using 4 sublayers.

The bidimensional frequency-dependent graphene complex conductivity $\sigma_{2D}(\omega)$ is modelled by the Kubo formula [7,8]:

$$\sigma_{2D}(\omega) = \frac{2e^2 k_b T}{\pi \hbar} \frac{i}{\omega + \frac{i}{\tau}} \log[2\cosh(E_F/k_b T)] + \frac{e^2}{4\hbar}\left[H\left(\frac{\omega}{2}\right) + \frac{4i\omega}{\pi}\int_0^{+\infty} d\varepsilon \frac{H(\varepsilon) - H(\frac{\omega}{2})}{\omega^2 - 4\varepsilon^2}\right]$$

where $\omega$ is the angular frequency, $e$ the elementary electric charge, $k_b$ Boltzmann's constant, $T$ the temperature, $\tau$ the finite electronic relaxation time, $E_F$ the Fermi energy, and $H(\varepsilon)$ is expressed as

$$H(\varepsilon) = \frac{\sinh(\hbar\varepsilon/k_b T)}{\cosh(E_F/k_b T) + \cosh(\hbar\varepsilon/k_b T)}$$

is a convenient numerical form avoiding singularities in the expression including a difference of Fermi distributions. This conductivity expression takes into consideration both intraband electron-photon scattering process and direct interband transitions across the optical gap. The latter can be ignored if the incident energy $\hbar\omega$ is smaller than $E_F$; consequently, the intraband conductivity takes the following Drude expression form at zero temperature:

$$\sigma_{2D}(\omega) = \frac{e^2}{\pi\hbar^2}\frac{iE_F}{\omega + \frac{i}{\tau}}.$$

The graphene complex permittivity $\varepsilon(\omega)$ is related to the bidimensional conductivity

$$\varepsilon(\omega) = \varepsilon_b + i\frac{\sigma_{2D}}{\varepsilon_0 \omega d_{graph}}$$

with $d_{graph} = 0.34$ nm corresponding to the interlayer distance in graphite. Actually, graphene is considered here as a thin conducting layer of finite thickness. $E_F$ can be estimated to 0.7 eV, i.e. a doping carrier density n ≈ $3.6 \; 10^{11}$ cm$^{-2}$ [1], the electronic relaxation time is set to $\tau = 1 \; 10^{-13}$s and the temperature T = 300 K.

- **S7: Approximations**

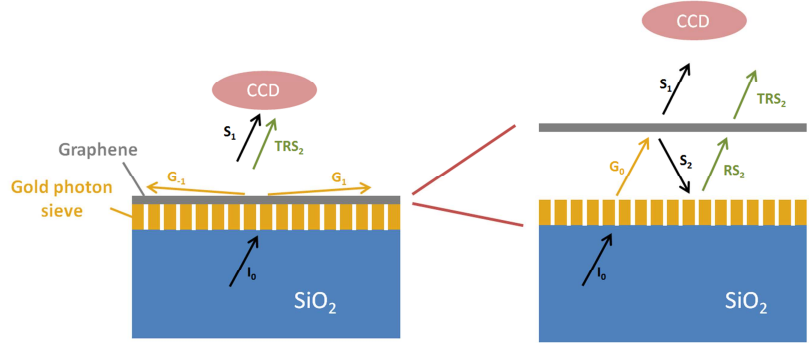

Figure S4: Explanation of the different diffraction order captured by the CCD camera in SHG experiment

In the semi-analytical method presented here, only the specular diffraction order is used (the specular terms S1 and S2 on figure S4). We proceed in this way because this is the sole diffraction order recorded by the experiment due to the numerical aperture of the whole detector (±6.25°). Other diffraction orders are either evanescent or emitted with angles greater than 20° (depending on the incident azimuthal angle) compared to the normal to the surface of the photon sieve (this can be easily checked by looking the wave vector $K = 2k_\parallel + g + \sqrt{4k^2 - |2k_\parallel + g|^2}e_z$ of a diffracted order g related to SHG). Furthermore, by symmetry argument, S1 = S2. The recorded intensity in the CCD camera can be written as:

$$S_1 + S_2 R_{gold} T_{graphene} \approx 1.6 S_1 < 2 S_1$$

where $R_{gold} = 0.63$ is the reflection coefficient at the gold photon sieve interface at the frequency 2ω and $T_{graphene} = 0.98$ is the transmission coefficient of graphene at the frequency 2ω. Consequently, the recorded intensity at the CCD camera is majored by 2S1.

By doing this approximation, the exact components are underestimated by about 10% ($\chi^{(2)} \propto \sqrt{\frac{0.8\, I(2\omega)}{I^2(\omega)}}$). We are not seeking a metrological precision for those values used for further calculations, the order of magnitude are already important. Moreover, a difference of three orders of magnitude is found between the $|d_{31} + d_{33}|$ and $|d_{15}|$ element tensors. This difference is prominent.

If we consider the role of the uncertainties due to the gold/glass interface, due to the correction of the near field from graphene, and due to our simplified scattering model, then the uncertainties on the values of the elements of $\chi^{(2)}$ are about 10%. Now, if we consider the effective scalar $\chi^{(2)}_{eff}$, the standard derivation from our experimental measurements leads to $\chi^{(2)}_{eff} = 5.7\ (\pm 0.8) \times 10^5$ pm²/V at 95% confidence level., i.e. 14% of relative uncertainty.

- **S8: References**